\newcommand{\rem}[1]{}
\newcommand{\beq}{\begin{equation}}
\newcommand{\eeq}{\end{equation}}
\newcommand{\beqa}{\begin{eqnarray}}
\newcommand{\eeqa}{\end{eqnarray}}
\newcommand{\ba}{\begin{array}}
\newcommand{\ea}{\end{array}}

\documentclass[pra,twocolumn,showpacs,preprintnumbers,
amsmath,amssymb,floatfix]{revtex4}
\usepackage{hyperref}
\usepackage{epsfig}

\begin{document}

\title{Pair condensation in the BCS-BEC crossover \\ 
of ultracold atoms loaded onto a 2D square lattice} 

\author{Luca Salasnich and Flavio Toigo}
\affiliation{Dipartimento di Fisica e Astronomia ``Galileo Galilei'' and 
CNISM, Universit\`a di Padova, Via Marzolo 8, 35131 Padova, Italy}

\date{\today}

\begin{abstract}
We investigate the crossover from the Bardeen-Cooper-Schrieffer (BCS) state 
of weakly-bound Cooper pairs to the Bose-Einstein Condensate (BEC) 
of strongly-bound molecular dimers in a gas of ultracold atoms 
loaded on a two-dimensional optical lattice. By using the 
the mean-field BCS equations of the emerging Hubbard model and the 
concept of off-diagonal-long-range-order for fermions we calculate 
analytically and numerically the pair binding energy, the energy gap and the 
condensate fraction of Cooper pairs as a function 
of interaction strength and filling fractor of atoms in the lattice 
at zero temperature.
\end{abstract}


\maketitle

\section{Introduction} 

Several experimental groups
\cite{greiner,regal,kinast,zwierlein,chin,ueda} 
have observed in ultracold alkali-metal atoms 
the predicted \cite{eagles,leggett,noziers} crossover 
from the Bardeen-Cooper-Schrieffer (BCS)
state of weakly bound Fermi pairs to the Bose-Einstein condensate (BEC)
of molecular dimers.
In two \cite{zwierlein,ueda} of these experiments  
the condensate fraction of Cooper pairs \cite{yang} 
has been studied with two hyperfine component Fermi vapours of $^6$Li atoms. 
The experimental data of the condensate fraction, which is directly related
to the off-diagonal-long-range order of the two-body density
matrix of fermions \cite{penrose,campbell}, are in quite good agreement
with mean-field theoretical predictions 
\cite{sala-odlro,ortiz} and Monte-Carlo 
simulations \cite{astrakharchik} at zero temperature, 
while at finite temperature beyond-mean-field 
corrections are needed \cite{ohashi}. 
Recently the condensate fraction in the BCS-BEC crossover 
has been theoretically investigated for a two-dimensional (2D) 
uniform Fermi gas \cite{sala-odlro2}, 
for a uniform three-spin-component Fermi gas with SU(3) 
symmetry \cite{sala-odlro3}, for a 2D uniform two-component 
Fermi gas with Rashba spin-orbit coupling \cite{sala-odlro4,cinesi}, 
and also for neutron matter \cite{sala-odlro5}.  
Two years ago 2D degenerate Fermi gases 
have been experimentally realized 
for ultra-cold atoms in a highly anisotropic 
disk-shaped potential \cite{turlapov}. 

Motivated by these recent theoretical 
and experimental achievements, in the present paper 
we analyze the condensate fraction in the BCS-BEC crossover 
for a quasi-2D two-component Fermi gas 
under optical confinement, which gives rise to a two-dimensional 
square lattice \cite{stoof}. 
In particular we study the energy gap and the condensate 
fraction of Cooper pairs 
as a function of the interaction strength (or equivalently 
as a function of binding energy of pairs) and filling factor 
of atoms in the lattice by using the concept of 
off-diagonal-long-range-order \cite{yang,penrose,campbell} and solving 
the mean-field BCS equations \cite{stoof}. 
The paper is organized as follows. In Section II we introduce 
the model Hamiltonian which describes two-spin-component Fermi 
atoms loaded onto a quasi-2D optical lattice. In Section III 
we discuss and solve the zero-temperature mean-field BCS 
equations as a function of the 
adimensional ratio between the interaction energy per site 
and the tunneling energy, calculating the binding energy 
of atomic pairs, the chemical potential and the energy gap 
order parameter. In particular, we compare the numerical 
results obtained by using the exact density of states with the 
analytical ones derived from an approximated density of states.  
In Section IV we calculate the condensate 
fraction of atomic pairs investigating 
the dependence of the condensate fraction on the relevant parameters 
of the system: scaled inter-atomic strength and filling factor. 
The paper is concluded by Section V. 

\section{Fermi atoms on a quasi-2D lattice} 

The Hamiltonian of a confined dilute and ultracold gas 
of two-component Fermi atoms is given by 
\beqa 
{\hat H} &=& \int d^3{\bf r} \sum_{\sigma} 
{\hat \psi}_{\sigma}^+({\bf r}) 
\left[ - {\hbar^2\over 2m}\nabla^2 + V_{ext}({\bf r}) 
\right] {\hat \psi}_{\sigma}({\bf r}) 
\nonumber
\\
&+& g  \int d^3{\bf r} \,  
{\hat \psi}_{\uparrow}^+({\bf r}) \, 
 {\hat \psi}^+_{\downarrow}({\bf r}) \,  
{\hat \psi}_{\downarrow}({\bf r}) \, 
{\hat \psi}_{\uparrow}({\bf r}) \; , 
\label{ham0} 
\eeqa 
where ${\hat \psi}_{\sigma}({\bf r})$ is the fermionic field operator 
that destroys an atom of pseudo-spin $\sigma$ ($\sigma=\uparrow,\downarrow$)  
at the position ${\bf r}$ and $g=4\pi \hbar^2a_s/m$ is the interaction 
strength of the contact inter-particle potential with $a_s$ the s-wave 
scattering length. The external optical potential 
\beq 
V_{ext}({\bf r}) = V_{lat}(x,y) + {1\over 2} m \omega_z^2 z^2 
\eeq
produces a harmonic confinement 
along the $z$ axis and a periodic potential 
\beq 
V_{lat}(x,y) = V_0 \, 
\left( \cos^2{({2\pi\over \lambda} x)} + \cos^2{({2\pi\over \lambda} y)}
\right) 
\;  
\eeq 
in the $(x,y)$ plane, with $\lambda$ the wavelength 
of the laser light which determines the optical lattice \cite{stoof}. 
The minima of the lattice potential form 
a two-dimensional square lattice with sites in the positions 
${\bf R}_{\bf i}=a \, {\bf i} = a \, (i_x,i_y)$, 
where $a=\lambda/2$ is the lattice spacing and 
${\bf i}=(i_x,i_y)$ is a 2D vector of integer numbers. 

Using the set of Wannier functions in the lowest Bloch band \cite{stoof}, 
where the Wannier function $W_{{\bf i}}(x,y)$ is maximally localized at 
site ${\bf R}_{\bf i}$, we can expand the fermionic field 
operator as: 
\beq 
{\hat \psi}_{\sigma}({\bf r}) = 
\sum_{\bf i} {\hat c}_{{\bf i}\sigma} \, 
W_{\bf i}(x,y) \, {e^{-z^2/(2a_z^2)}\over \pi^{1/4} a_z^{1/2}} \; ,   
\eeq
where ${\hat c}_{{\bf i}\sigma}$ and ${\hat c}_{{\bf i}\sigma}$ obey 
the usual Fermi anti-commutation relations, and 
$a_z=\sqrt{\hbar/(m\omega_z)}$ is the characteristic length 
of the strong harmonic confinement along the $z$ axis, which induces 
a quasi-2D confinement if $\hbar\omega_z$ is much larger than 
the other energies of the system. Under these conditions, 
the Hamiltonian (\ref{ham0}) can be written as 
\beq 
{\hat H} = -t \sum_{\langle {\bf i}{\bf j} \rangle \, \sigma} 
{\hat c}^+_{{\bf i}\sigma} {\hat c}_{{\bf j}\sigma} 
+ U \sum_{\bf i} {\hat n}_{{\bf i}\uparrow}{\hat n}_{{\bf i}\downarrow} \; , 
\label{ham} 
\eeq 
where  $\langle {\bf i}{\bf j} \rangle$ means nearest neighbor sites, 
\beq 
t = - \int dx\, dy \, W_{{\bf i}}^*(x,y) 
\left[  - {\hbar^2\over 2m}\nabla^2 + V_{latt}(x,y) 
\right] W_{{\bf j}}(x,y)
\eeq
is the hopping parameter ($t>0$), i.e. the tunneling energy 
between nearest neighbor sites, and 
\beq 
U = {g\over \pi a_z}  \int dx\, dy \, |W_{{\bf i}}(x,y)|^4 
\eeq
is the on-site strength of the 
inter-atomic interaction. ${\hat n}_{{\bf i}\sigma}= 
{\hat c}_{{\bf i}\sigma}^+{\hat c}_{{\bf i}\sigma}$ 
is the number operator which describes the number 
of atoms with spin $\sigma$ at the site ${\bf i}$, and consequently 
the total number operator reads 
\beq 
{\hat N}= \sum_{{\bf i} \, \sigma} \, {\hat n}_{{\bf i}\sigma} \; . 
\eeq
Notice that Eq. (7) holds under the conditions $|a_s|\ll a_z$ and $|a_s|\ll a$, 
which ensure the absence of confinement induced resonance \cite{cir} and 
no distorsion of Cooper pairs due to neighbor valleys of the 
optical confinement. 
In the Hubbard-like Hamiltonian (\ref{ham}) we have not included the 
tunneling energies between sites which are not nearest neighbor 
because they are exponentially suppressed. We have also 
assumed the on-site one-body energies to be the same on all sites 
and therfore dropped them as irrelevant \cite{stoof}. 

\section{Mean-field BCS equations}

It is well known that the BCS state appears only 
in the case of an attractive strength, i.e. $U<0$ \cite{stoof}. 
In the past the negative-$U$ Hubbard Hamiltonian has been investigated 
by various authors \cite{review} as a model for 
high-$T_c$ superconductivity. More recently, it has been used to study 
the BCS-BEC crossover on 2D and 3D lattices 
both at zero \cite{andrenacci,kujawa,capone} 
and finite temperature \cite{dupuis,tamaki}. 
As stressed in the introduction, motivated by recent 
theoretical and experimental achievement with ultracold atoms 
in optical lattices, here we reconsider the 2D negative-$U$ Hubbard 
Hamiltonian to investigate the pair condensation, and in particular 
the condensate fraction of Fermi atoms in the 2D 
lattice at zero temperature. Note that he condensate fraction has been 
calculated by Kujawa \cite{kujawa} in the 3D square lattice 
with a generalized Hubbard model, but only in the special case $|U|/t=\infty$ . 
In the following sections we calculate, as a function of $|U|/t$ 
and of the filling factor, the energy gap and 
condensate fraction in the 2D square lattice, analyzing also 
the pair binding energy, which is always finite 
in the 2D BCS-BEC crossover. 

We start by decoupling the interaction Hamiltonian of Eq. (\ref{ham}) in both 
normal and anomalous channels \cite{privitera}
\beqa 
{\hat n}_{{\bf i}\uparrow}{\hat n}_{{\bf i}\downarrow} &\simeq& 
\langle {\hat n}_{{\bf i}\uparrow}\rangle 
{\hat n}_{{\bf i}\downarrow} + 
{\hat n}_{{\bf i}\uparrow} \langle {\hat n}_{{\bf i}\downarrow}\rangle 
- \langle {\hat c}_{{\bf i}\uparrow}^+{\hat c}_{{\bf i}\downarrow}^+\rangle 
{\hat c}_{{\bf i}\uparrow}{\hat c}_{{\bf i}\downarrow}
\\
&-& {\hat c}_{{\bf i}\uparrow}^+{\hat c}_{{\bf i}\downarrow}^+  
\langle {\hat c}_{{\bf i}\uparrow}{\hat c}_{{\bf i}\downarrow} \rangle 
- \langle {\hat n}_{{\bf i}\uparrow}\rangle 
\langle {\hat n}_{{\bf i}\downarrow}\rangle 
+ 
\langle  {\hat c}_{{\bf i}\uparrow}^+{\hat c}_{{\bf i}\downarrow}^+\rangle   
\langle {\hat c}_{{\bf i}\uparrow}{\hat c}_{{\bf i}\downarrow} \rangle \; . 
\nonumber
\eeqa
We also assume 
\beq 
{n\over 2} = 
\langle {\hat n}_{{\bf i}\uparrow}\rangle = 
\langle {\hat n}_{{\bf i}\uparrow}\rangle 
\eeq
and introduce the (real) mean-field, site-independent,  
gap order parameter 
\beq 
\Delta = - U 
\langle  {\hat c}_{{\bf i}\uparrow}^+{\hat c}_{{\bf i}\downarrow}^+\rangle  
= - U \langle  {\hat c}_{{\bf i}\downarrow}{\hat c}_{{\bf i}\uparrow}\rangle \; . 
\eeq 
In this way we obtain the mean-field Hamiltonian 
\beqa 
{\hat H}_{MF} &=& -t \sum_{\langle {\bf i}{\bf j} \rangle \, \sigma} 
{\hat c}^+_{{\bf i}\sigma} {\hat c}_{{\bf j}\sigma} 
+ {U n\over 2}\sum_{\bf i} \left( {\hat n}_{{\bf i}\uparrow} 
+ {\hat n}_{{\bf i}\downarrow} \right) 
\\
&+& \Delta \sum_{\bf i} 
\left(  {\hat c}_{{\bf i}\uparrow}{\hat c}_{{\bf i}\downarrow} 
+ {\hat c}_{{\bf i}\downarrow}^+{\hat c}_{{\bf i}\uparrow}^+ \right) 
- {Un^2\over 4} N_s + {\Delta^2\over U}N_s \; , 
\nonumber
\label{ham-mf}  
\eeqa
where $N_s$ is the number of lattice sites. 

In the dual space of wavevectors ${\bf k}=(k_x,k_y)$,  
setting 
\beq 
{\hat c}_{{\bf i}\sigma} = \sum_{\bf k} {\hat c}_{{\bf k}\sigma} 
\ {e^{i{\bf k}\cdot {\bf R}_{\bf i}}\over \sqrt{N_s}} \; , 
\eeq
where ${\hat c}_{{\bf k}\sigma}$ destroys an atom of 
spin $\sigma$ and wavevector ${\bf k}$, 
the mean-field Hamiltonian (\ref{ham-mf}) becomes 
\beqa 
{\hat H}_{MF} &=& \sum_{\bf k} 
\left( \epsilon_{\bf k} + {Un\over 2} \right) 
{\hat c}_{{\bf k}\sigma}^+ 
{\hat c}_{{\bf k}\sigma} 
\\
&+& \Delta \sum_{{\bf k}} \left( 
{\hat c}_{{\bf k}\uparrow} 
{\hat c}_{-{\bf k}\downarrow}
+ {\hat c}_{-{\bf k}\downarrow}^+
{\hat c}_{{\bf k}\uparrow}^+ \right) 
- {Un^2\over 4} N_s + {\Delta^2\over U}N_s \; , 
\nonumber
\eeqa
where 
\beq 
\epsilon_{\bf k} = - 2 t 
\left( \cos{(k_x a)} + \cos{(k_y a)} \right) \; , 
\label{single}
\eeq
is the single-particle energy. 
We stress that we are considering only the lowest Bloch band. 
This single-band approximation for the BCS-BEC crossover is 
reliable since the crossover occours at magnetic fields that are 
relatively far away from the Feshbach resonance underlying it 
\cite{stoof2}. Moreover the approximation is reliable under 
the following conditions \cite{stoof2,georges}: 
i) there are no more than two fermions per site; 
ii) the two lowest bands do not overlap, implying that 
$V_0\gg E_r$, which means $8t\ll E_r$, 
and $|U| \ll E_g$. $E_r=\hbar^2k_L^2/(2m)$ is the 
recoil energy with $k_L=2\pi/a$ the wavevector of the 2D optical 
lattice, and $E_g$ is the energy gap between the first and 
the second Bloch band.

We calculate the thermodynamic potential
\beq
\Omega = \langle {\hat H}_{MF}\rangle - \mu \, \langle \hat{N} \rangle  \; ,
\label{ther}
\eeq
where $\mu$ is the chemical potential which determines the
average number $N=\langle{\hat N}\rangle$ of fermions, 
by introducing the Bogoliubov canonical transformation:  
\beq 
{\hat \alpha}_{\bf k} = u_{\bf k} \, {\hat c}_{{\bf k}\uparrow} 
-  v_{\bf k} \, {\hat c}_{-{\bf k}\downarrow}^+ \; , 
\quad\quad 
{\hat \beta}_{\bf k} = u_{\bf k} \, {\hat c}_{-{\bf k}\downarrow} 
+  v_{\bf k} \, {\hat c}_{{\bf k}\uparrow}^+ \; , 
\eeq
where $u_{\bf k}$ and $v_{\bf k}$ are real and 
$u_{\bf k}^2+v_{\bf k}^2=1$. After the mimimization 
of $\Omega$ with respect to $\mu$ and $\Delta$ 
we recover the standard BCS equation \cite{stoof,privitera}
for the average number of particles per site 
\beq
n = 2 {1 \over N_s} \sum_{\bf k} v_{\bf k}^2  \; ,
\label{bcs1}
\eeq
and the familiar BCS gap equation
\beq
{1\over |U|} = {1 \over N_s} \sum_{\bf k} {1\over 2 E_{\bf k}} \; ,   
\label{bcsGap}
\eeq
where the quasi-particle 
amplitudes $u_{\bf k}$ and $v_{\bf k}$ are given by 
\beq 
\label{vk}
v_{\bf k}^2 = {1\over 2} \left( 1 - \frac{\epsilon_{\bf k} 
- h }{E_{\bf k}} \right) \, ,  
\eeq 
and $u_{\bf k}^2=1 - v_{\bf k}^2$. Here the Bogoliubov energy reads 
\beq
E_{\bf k}=\left[\left(\epsilon_{\bf k}-h \right)^2 
+ \Delta^2\right]^{1/2} 
\eeq
where 
\beq 
h = \mu - {Un\over 2} \; , 
\eeq
is the effective chemical potential which takes into account 
the Hartree interaction. 
The effective chemical potential $h$ and the gap energy $\Delta$ 
are obtained by solving equations (\ref{bcs1}) and (\ref{bcsGap}). 
In the continuum limit $\sum_{\bf k}\to a^2 N_s \int_{\cal BZ} 
d^2{\bf k}/(2\pi)^2$ 
and introducing the density of states (DOS) per site 
\beqa 
{\cal D}(\epsilon) &=& a^2 \int_{\cal BZ} {d^2{\bf k}\over (2\pi)^2} \; 
\delta(\epsilon_{\bf k}-\epsilon) 
\label{DoS-exact}
\\
&=& {1\over 2\pi^2 t} 
K\left( \sqrt{1-{\epsilon^2\over 16t^2}} \right) \, 
\Theta\left(1-{\epsilon^2\over 16t^2}\right) \; , 
\nonumber 
\eeqa
where ${\cal BZ}=[-\pi/2,\pi/a]\times [-\pi/a,\pi/2]$ is the first 
Brillouin zone, $K(x)$ is the complete Elliptic integral of the first kind 
and $\Theta(x)$ is the step function, 
the number equation (\ref{bcs1}) and the gap equation  (\ref{bcsGap}) 
can be written as 
\beqa
n &=& \int_{-4t}^{4t} d\epsilon \, {\cal D}(\epsilon) \,  
\left(1 - {\epsilon-h\over \sqrt{(\epsilon-h)^2+\Delta^2}}\right) \; , 
\label{exact1}
\\
{1\over |U|} &=&  \int_{-4t}^{4t} d\epsilon \, {\cal D}(\epsilon) \, 
{1\over 2\sqrt{(\epsilon-h)^2+\Delta^2}} \; .  
\label{exact2}
\eeqa

\begin{figure}
\centerline{\epsfig{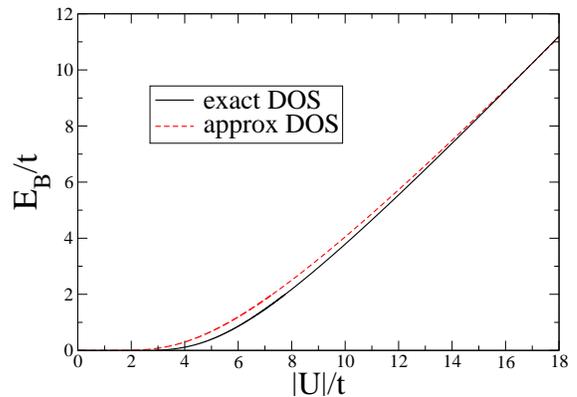}}
\caption{(Color online). Scaled binding energy $E_B/t$ as a function
of the scaled interaction strength $|U|/t$, with $t$ the
tunneling energy. Solid lines are the results obtained with the 
exact density of states (exact DOS) given by Eq. (\ref{DoS-exact}), while 
dashed lines are the results obtained with the approximate
density of states (approx DOS) given by Eq. (\ref{DoS-approx}).}
\label{fig1}
\end{figure}

As discussed in \cite{review}, quite generally in two dimensions 
a bound-state energy $E_B$ exists for any value 
of the negative interaction strength $U$. For the contact potential 
the bound-state equation in the lattice is 
\beq 
{1 \over |U|} =  \int_{-4t}^{4t} d\epsilon \, {\cal D}(\epsilon) \, 
{1\over 2(\epsilon - \epsilon_{\bf 0}) + E_B} \, ,  
\label{binding}
\eeq 
where $\epsilon_{\bf 0}=-4t$ is the lower value of the single-particle 
energy $\epsilon_{\bf k}$, occurring at ${\bf k}={\bf 0}$. 
If we approximate the true DOS with a constant value 
in the interval $[-4t,4t]$, i.e. 
\beq 
{\cal D}(\epsilon)\simeq {1\over 8t} \, 
\Theta\left(1-{\epsilon^2\over 16t^2}\right) \; ,
\label{DoS-approx}
\eeq 
that ensures the normalization 
\beq 
\int_{-4t}^{4t} d\epsilon \, {\cal D}(\epsilon) =1 \, ,
\eeq 
the bound-state equation can be solved analytically giving   
\beq 
{1 \over |U|} = {1\over 16 t} 
\ln{\left|{E_B+16t\over E_B}\right|} \; . 
\eeq

In Fig. \ref{fig1} we plot the 
binding energy $E_B/t$ as a function
of the interaction strength $|U|$ obtained with this approximate 
formula (dashed line). For comparison we plot also the 
exact result (solid line), obtained by numerically 
solving Eq. (\ref{binding}). The figure shows that the agreement between 
the two curves is extremely good. The BCS-BEC crossover is governed 
by the adimensional parameter $|U|/t$ or equivalently 
by the scaled binding energy $E_B/t$. The limit of large 
tunneling and small interaction $|U|/t\ll 1$ corresponds 
to the BCS regime where $E_B/t$ is close to zero. 
Instead the limit of strong 
localization and large interaction $|U|/t\gg 1$ 
corresponds to the BEC regime where $E_B/t$ is large. 

\begin{figure}
\centerline{\epsfig{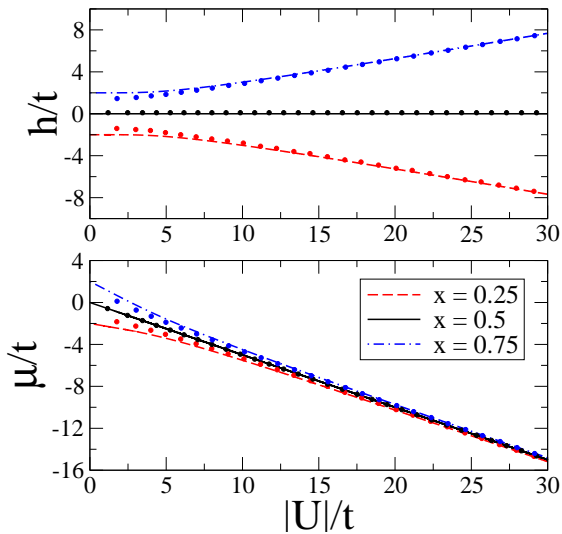}}
\caption{(Color online). Scaled effective chemical potential $h/t$ (upper 
panel) and scaled chemical potential $\mu/t$ (lower panel) as a function
of the scaled interaction strength $|U|/t$, with $t$ the
tunneling energy. Results obtained 
for three values of the filling factor $x=n/2$.
Filled circles are the results obtained with the 
exact density of states 
given by Eq. (\ref{DoS-exact}), while lines are the results 
obtained with the approximate
density of states given by Eq. (\ref{DoS-approx}).}
\label{fig2}
\end{figure}

The quite good agreement between the solid curve and the dashed curve 
of Fig. \ref{fig1} suggests that one could 
use the approximate 
DOS to study various ground-state properties of the system in the BCS-BEC. 
Within the approximation of a constant DOS in the band, 
i.e. Eq. (\ref{DoS-approx}), 
the number density equation and the gap equation read 
\beqa
n &=& {1\over 8t} 
\Big( 8t - \sqrt{(4t-h)^2+\Delta^2} 
\label{approx1}
\\
&+& \sqrt{(4t+h)^2+\Delta^2} \Big) \; ,
\nonumber 
\\
{1\over |U|} &=&  {1\over 16t} \, 
\ln{\left|{h+\sqrt{h^2+\Delta^2}\over h-8t+
\sqrt{(h-8t)^2+\Delta^2}}\right|} \; .
\label{approx2}
\eeqa
It is then straightforward to plot (see Fig. \ref{fig2})  
the effective chemical potential $h$ 
(upper panel) and the chemical potential $\mu$ (lower panel) 
as a function of the scaled 
interaction strength $|U|/t$, 
for different values of the filling factor $x=n/2$ 
($0\leq x \leq 1$). In the figure the lines 
are obtained by using Eqs. (\ref{approx1}) and (\ref{approx1}) 
based on the approximate DOS of Eq. (\ref{DoS-approx}), while 
the filled circles are obtained by using Eqs. (\ref{exact1}) 
and (\ref{exact2}) with the exact DOS of Eq. (\ref{DoS-exact}). 

Fig. \ref{fig2} shows that 
at half filling ($x=0.5$) the 
effective chemical potential $h$ remains always 
constant and equal to zero,  
and the corresponding chemical potential $\mu$ follows the 
simple law $\mu =-|U|/2$. Moreover, the lower panel of Fig. \ref{fig2}
shows that, at fixed filling factor $x$, the chemical 
potential $\mu$ as a function 
of $U$ is close to a straight line (it is true straight line 
only for $x=0.5$) and approaches $\mu\simeq -|U|/2$ for large $|U|$. 

\begin{figure}
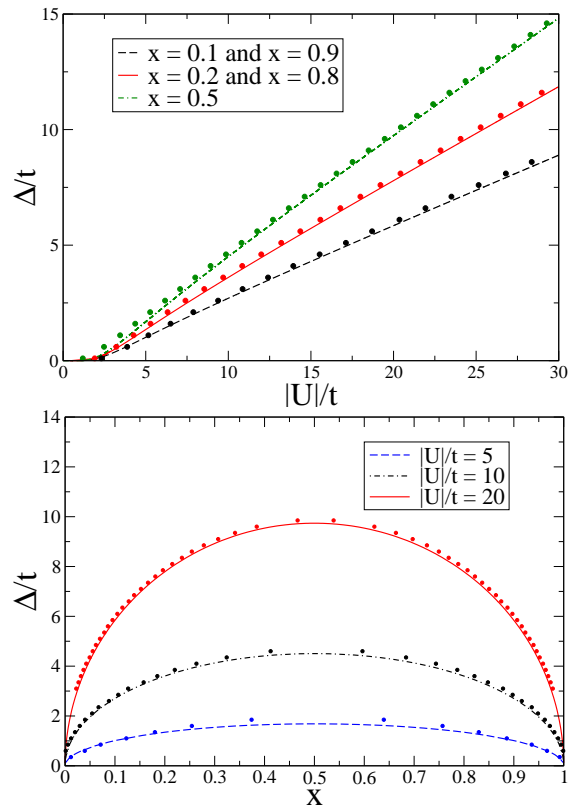

\centerline{\epsfig{file=delta.eps,width=7.4cm,clip=}}
\centerline{\epsfig{file=chi-delta.eps,width=7.4cm,clip=}}
\caption{(Color online). Upper panel: Scaled energy gap $\Delta/t$ 
as a function of scaled interaction strength $|U|/t$
with $t$ the tunneling energy. The three curves correspond 
to five different values of the filling factor $x=n/2$.
Lower panel: Scaled energy gap $\Delta/t$ 
as a function of filling factor $x=n/2$, where 
the three curves correspond to three different values of the 
scaled interaction strength $|U|/t$, with $t$ the tunneling energy.
Filled circles are the results obtained with the 
exact density of states given by Eq. (\ref{DoS-exact}), 
while lines are the results obtained with the approximate
density of states given by Eq. (\ref{DoS-approx}).} 
\label{fig3}
\end{figure} 

In Fig. \ref{fig3} we plot the energy gap $\Delta$ vs 
interaction strength $|U|$ (upper panel) and vs 
filling factor $x$ (lower panel). 
The upper panel shows that, at fixed filling factor $x$,  
the energy gap $\Delta$ grows by increasing the 
scaled interaction strength $|U|/t$, 
that is by increasing the localization. Instead, the lower 
panel shows that, at fixed 
scaled interaction strength $|U|/t$, the scaled energy gap 
$\Delta/t$ reaches its maximum value at half filling $x=1/2$, 
i.e. when on the average there is one atom per site. 
This effect is clearly seen in the lower panel of Fig. \ref{fig3} 
where we consider three values of $|U|/t$. Notice 
that the behavior of $\Delta$ as a function of $x$ is perfectly 
symmetric with respect to $x=1/2$ (half filling). 
Also in Fig. \ref{fig3} the agreement between the results 
obtained with the exact DOS and the ones calculated with the 
approximate DOS is quite good, and it improves by increasing $|U|/t$. 
Motivated by this finding, in the remaining part of the paper we use 
the approximate DOS, which is much simpler for numerical 
computations and produces analytical results. 

\section{Condensate fraction}

The main task of the paper is to analyze the condensate 
fraction of fermions. 
As shown by Yang \cite{yang}, the BCS state 
guarantees the off-diagonal-long-range-order \cite{penrose} 
of the Fermi gas, namely that, in the limit wherein
both unprimed coordinates approach an infinite distance from the primed 
coordinates, the two-body density matrix factorizes as follows: 
\beqa
\label{2bodydm}
\langle 
{\hat \psi}^+_{\uparrow}({\bf r}_1') 
{\hat \psi}^+_{\downarrow}({\bf r}_2') 
{\hat \psi}_{\downarrow}({\bf r}_1) 
{\hat \psi}_{\uparrow}({\bf r}_2) 
\rangle 
\\
= \langle 
{\hat \psi}^+_{\uparrow}({\bf r}_1') 
{\hat \psi}^+_{\downarrow}({\bf r}_2') 
\rangle \langle 
{\hat \psi}_{\downarrow}({\bf r}_1) 
{\hat \psi}_{\uparrow}({\bf r}_2)
\rangle \, . 
\nonumber 
\eeqa
The largest eigenvalue of the two-body density matrix (\ref{2bodydm})
gives the number of pairs in the lowest two-particle state, 
i.e. the condensate number of Fermi pairs \cite{leggett,yang,campbell}. 
In this way, the number $N_0$ of condensed fermions is given by 
\beq
\label{ODLRO:def}
N_0 = 2 \int d^3{\bf r} \, d^3{\bf r}' \; | \langle
{\hat \psi}_{\downarrow}({\bf r})
{\hat \psi}_{\uparrow}({\bf r}')  \rangle |^2 
= 2 \sum_{{\bf i}{\bf j}} \; | \langle 
{\hat c}_{{\bf i}\downarrow} 
{\hat c}_{{\bf j}\uparrow}\rangle |^2 .  
\eeq 
Notice that, as said above,  $N_0$ counts the number of 
condensed fermions: $0\leq N_0\leq N$ 
\cite{sala-odlro3} and not of condensed pairs. 
It is then quite easy to show that the the condensate 
number of atoms per site is 
\beq 
n_0 = 2 {1\over N_s} \sum_{\bf k} u_{\bf k}^2 v_{\bf k}^2 \; . 
\eeq 
With the help of Eqs. (\ref{vk}) this number  is thus given by :
\beq 
n_0 = {\Delta^2\over 2} \int_{-4t}^{4t} d\epsilon \, {\cal D}(\epsilon) \, 
{1\over (\epsilon-h)^2+\Delta^2} \; ,   
\eeq
and using the approximate DOS of Eq. (\ref{DoS-approx}) it reads: 
\beq 
n_0 = {\Delta\over 16 t} 
\left( \arctan{\left({4t-h\over \Delta}\right)} +  
\arctan{\left({4t+h\over \Delta}\right)} \right) \; . 
\label{approx3}
\eeq

\begin{figure}
\centerline{\epsfig{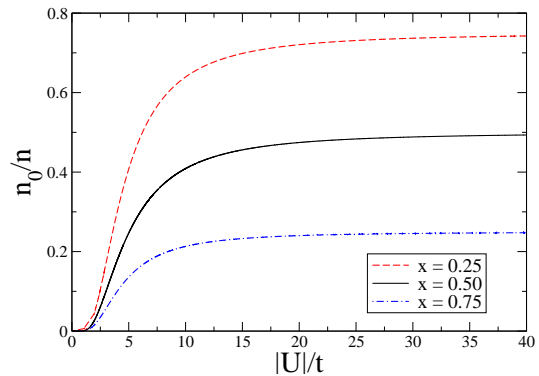}}
\caption{(Color online). Condensate fraction $n_0/n$ 
as a function of scaled interaction strength $|U|/t$
with $t$ the tunneling energy. The three curves correspond 
to three different values of the filling factor $x=n/2$.} 
\label{fig4}
\end{figure} 

Fig. \ref{fig4} shows the condensate fraction $n_0/n$ of fermions, 
calculated with Eqs. (\ref{approx1}), (\ref{approx2}) and (\ref{approx3}),   
as a function of scaled interaction strength $|U|/t$ 
for three values of the filling factor $x$. We have verified 
that the plotted results are in good agreement with the 
ones obtained by using the exact DOS, 
except in the case of very small values of $|U|/t$. In any case, 
the condensate fraction $n_0/n$ vanishes when 
the scaled interaction strength $|U|/t$ goes to zero. Moreover, 
as shown in the figure,  the condensed fraction 
grows very fast for  values of the scaled 
interaction strength $|U|/t\leq 8$, it shows a shoulder, 
and then it reaches its  
asymptotic value  $n_0/n \simeq 1-x$ rather slowly. 

\begin{figure}
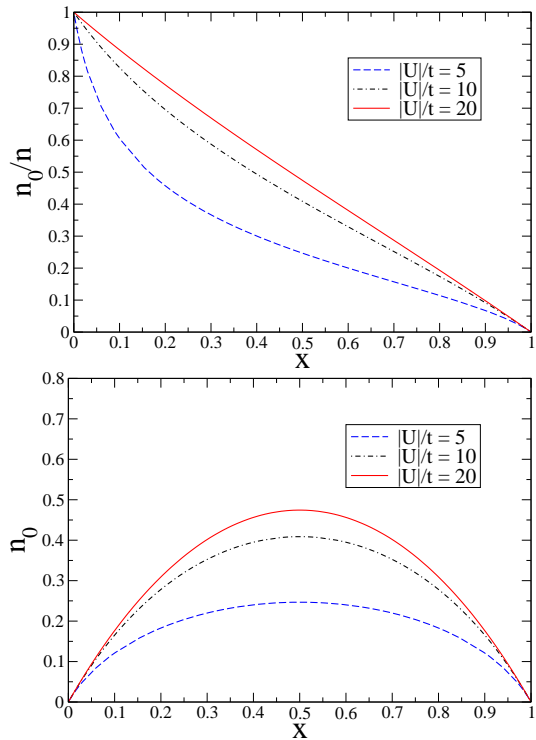

\vskip 0.4cm
\centerline{\epsfig{file=chi-cond.eps,width=7.cm,clip=}}
\centerline{\epsfig{file=chi-n0.eps,width=7.cm,clip=}}
\caption{(Color online). Upper panel: Condensate fraction $n_0/n$ 
as a function of the filling factor $x=n/2$. 
Lower panel: Number $n_0$ of condensed atoms per site. 
The curves correspond to three different values of the 
scaled interaction strength $|U|/t$, 
with $t$ the tunneling energy.} 
\label{fig5}
\end{figure} 

This result is confirmed in the upper panel of Fig. \ref{fig5}, 
where we report the condensate fraction $n_0/n$ 
as a function of the filling factor $x$ at fixed 
scaled interaction strength $|U|/t$.  
The figure clearly shows that $n_0/n$ ranges from one 
to zero, being extremely close to one for $x \ll 1$ and 
approaching zero as $x$ goes to $1$. 
This means that there is a full BEC-BCS crossover by increasing $x$ 
at constant scaled interaction strength $|U|/t$. 
Moreover, if the scaled interaction strength $|U|/t$ is large, 
the condensate fraction $n_o/n$ follows a straight line 
during the BEC-BCS crossover. 
For the sake of completeness, in the lower panel of Fig. \ref{fig5} 
we plot also the number $n_0$ of condensed atoms per site 
as a function of the filling factor $x$. 
The results show that the curves of $n_o$ vs $x$ have a behavior 
similar to those of $\Delta$ vs $x$ (see Fig. \ref{fig3}). 
In the limit $|U|/t\to \infty$ one finds that $n_0=(1-x)2x$ 
and consequently $n_0/n=(1-x)$. 

Finally, we observe that, after a simple rescaling of the chemical 
potential, namely ${\tilde h}=h+4t$, in the limit $t\to \infty$ 
with $ta^2\to \pi\hbar^2/m$, Eq. (\ref{approx3}) becomes 
the condensate number equation found in Ref. \cite{sala-odlro2} 
for the 2D uniform superfluid Fermi gas. 

\section{Conclusions} 

By using the mean-field extended BCS theory and 
the concept of off-diagonal long-range order, that is 
the existence of a macroscopic eigenvalue of the two-body 
density matrix, we have investigated the condensate fraction 
of fermionic pairs in a uniform 2D Fermi gas. 
We have shown that the condensate number $n_0$ of fermi atoms per site 
is extremely useful 
to characterize the BCS-BEC crossover, that is induced by changing the 
adimensional ratio $|U|/t$ between the interaction energy $|U|$ 
and the tunneling energy $t$. In particular, we have found that 
both the scaled binding energy $E_B/t$ 
of atomic pairs and the condensate fraction $n_0/n$ 
grow by increasing the ratio $|U|/t$ at fixed filling 
factor $x=n/2$ (with $n$ the average number of fermions per site). 
In addition, our results suggest that 
fixing the ratio $|U|/t$, or equivalently the scaled 
binding energy $E_B/t$, there is a full BEC-BCS crossover 
by increasing the filling factor from zero to one. 
Finally, we have found that the analytical results obtained 
by using an approximate density of states are in quite 
good agreement with the numerical ones deduced from the exact density 
of states. In our calculations we have used the mean-field theory and 
it is important to stress that recent 
Monte Carlo simulations have shown that, at zero-temperature, 
beyond-mean-field effects
are negligible in the BCS side of the BCS-BEC crossover while they become
relevant in the deep BEC side \cite{astrakharchik,bertaina}. 
In any case, we think that our mean-field results, and the 
reliable analytical formulas we have obtained, can be of interest 
for near future experiments with degenerate gases 
made of alkali-metal atoms confined in quasi-2D optical lattices. 

In this paper we have investigated zero-temperature pair condensation. 
According to the Mermin-Wagner theorem \cite{stoof}, for an infinite 
2D system there is condensation (off-diagonal-long-range order) only 
at zero temperature. However, for a finite 2D system 
condensation could be possible also at non-zero temperature. 
The investigation of this issue, which requires a beyond mean-field approach,
for 2D fermions in a lattice is in progress. 
Another puzzling issue is the filling of the second Bloch band: 
we plan to investigate its effect on pair condensation 
by analyzing a multi-band version of the present theory.

\end{document}